\documentclass[letterpaper,twocolumn,showpacs,prl]{revtex4}

\newcommand{\nodagger}{\ensuremath\vphantom{\dagger}}

\newcommand{\be}{\begin{equation}}
\newcommand{\ee}{\end{equation}}
\newcommand{\bea}{\begin{eqnarray}}
\newcommand{\eea}{\end{eqnarray}}
\newcommand{\la}{\langle}
\newcommand{\ra}{\rangle}

\newcommand{\lp}{\left(}
\newcommand{\rp}{\right)}

\newcommand{\p}{\partial}

\renewcommand{\L}{{\cal L}}

\usepackage{amsmath}
\usepackage{graphicx}

\begin{document}
\title{Minimal excitation states of electrons in one-dimensional wires}
\author{J.~Keeling$^1$}
\author{I.~Klich$^2$}
\author{L.~S.~Levitov$^1$}

\affiliation{$^1$Department of Physics, Massachusetts Institute of
  Technology, 77 Massachusetts Ave, Cambridge, Massachusetts 02139, USA\\
$^2$Department of Physics, California Institute of
  Technology, Pasadena, CA 91125}

\begin{abstract}
  A strategy is proposed to excite particles from a Fermi sea in a
  noise-free fashion by electromagnetic pulses with realistic
  parameters.
  We show that by using quantized pulses of simple form one can
  suppress the particle-hole pairs which are created by a generic
  excitation.
  The resulting many-body states are characterized by one or several
  particles excited above the Fermi surface accompanied by no
  disturbance below it.
  These excitations carry charge which is integer for noninteracting
  electron gas and fractional for Luttinger liquid.
  The operator algebra describing these excitations is derived, and a
  method of their detection which relies on noise measurement is
  proposed.
\end{abstract}

\maketitle


Controlling 
single electrons is one of the
main avenues of research in nanoelectronics.
Once advanced far enough, it will bring about a range of
quantum-coherent single particle sources with full control over the
orbital and spin degrees of freedom.
Currently the efforts are mostly 
focused on employing localized electron states, 
trapped on metal islands~\cite{Delsing89,Geerligs90} 
or quantum dots~\cite{Kouwenhoven91}
and shuttled between
the dots or islands by electric pulses~\cite{Delsing89,Geerligs90,Kouwenhoven91} or acoustic fields~\cite{Talyanskii}.
It is of interest, however, to extend the concept
of single particle sources to the situation when electrons propagate
freely as part of a degenerate Fermi system.
If proved feasible, it would allow one to harness particle dynamics,
characterized by high Fermi velocity, $v_{\mathrm{F}}\sim 10^8\,{\rm cm/s}$, to
transmit quantum states in a solid and,
at low temperature, 
to use Fermi-Dirac statistics for generating many-particle 
entangled states\cite{Beenakker05,Samuelsson03,Beenakker03,Beenakker05a,Lebedev05}.

In this article we propose a scheme which allows the creation of
``clean" electric current pulses, free of particle-hole excitations.
We consider a 1d electron gas, serving as a prototype 
for carbon nanotube, quantum wire and point contact systems,
in which current is driven by voltage
pulses with a typical frequency small compared to the Fermi energy.
In this quasistationary regime
the electric response is described as
$I(t)=g_0V(t)$
with $g_0=e^2/h$ the Landauer conductance.
A current pulse, which carries total charge
$\Delta q=g_0\int V(t)dt$, is a collective many-body state
involving a number of fermions excited to a higher 
energy~\cite{Rychkov05}.
Microscopically, 
such a current pulse is described 
by a number of particle-hole excitations, with energies
of the order $\hbar/\tau$, where $\tau$ is the duration of the pulse.
As discussed in Refs.\cite{Rychkov05,Polianski02,Lesovik94} and below, these excitations 
can be probed by noise measurement~\cite{Schoelkopf98,Reydellet03}.

Here we show that, quite strikingly, by engineering the pulse profile
one can inhibit the particle-hole excitations.
We analyze the particle-hole content of current pulses
in a single-channel conductor,
and pose and solve the problem of minimizing the number of such excitations. 
The condition 
required for the excitation number to be
small is area quantization, $\int Vdt=nh/e$ ,
where $n$ is an integer. We show that such 
pulses, carrying integer charge 
$q= ne$, are accompanied by fewer excitations
than non-integer pulses.
Also, we address the question of optimizing the $V(t)$ profile,
and show that one can 
design pulses
which are totally free of particle-hole excitations.  
Such pulses excite $n$ electrons above the Fermi level, with other
electrons conspiring to fill the void and produce a complete Fermi
sea, with no holes.

\begin{figure}[tpb]
  \centering
  \includegraphics[width=3.5in]{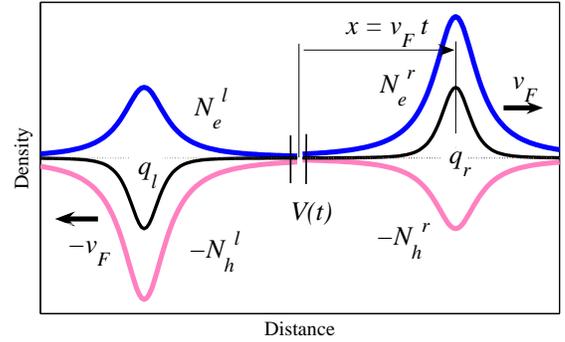}
\vspace{-8mm}
  \caption{Real space picture of counter-propagating
    electron and hole pulses, for a general time dependent field.
    In the special case of Eq.~(\ref{eq:phase}), $N_h^r=N_e^l=0$,
    $q_r=N_e^r$, $q_l=-N_h^l$.}
  \label{fig:real-space}
\vspace{-3mm}
\end{figure}

The properties of such ``ideal'' pulses may be inferred by asking for 
$V(t)$
that creates the minimum number of excitations 
moving to the right or to the left
\begin{equation}
  \label{eq:charge-nex}
  N_{\mathrm{ex}}^\alpha  = N_{\mathrm{e}}^\alpha  + N_{\mathrm{h}}^\alpha 
  ,\quad
  N_{\mathrm{e}}^\alpha =\sum_{\epsilon>\epsilon_{\mathrm{F}}}
  \langle a^{\dagger}_\epsilon a_\epsilon \rangle
  ,\
  N_{\mathrm{h}}^\alpha =\sum_{\epsilon<\epsilon_{\mathrm{F}}}
  \langle a_\epsilon a^{\dagger}_\epsilon \rangle
  ,
\end{equation}
where $\alpha=r,\,l$ for right and left movers, 
and $N_{\mathrm{e}}^\alpha $, $N_{\mathrm{h}}^\alpha $ are the numbers of excited 
electrons and holes
(Fig.\ref{fig:real-space}). 
The operator $a^{\dagger}_\epsilon$ creates a particle in a single
particle eigenstate, labelled by the energy of that state, the sum is
restricted to right or left moving particles, $k\sim k_{\mathrm{F}},\,-k_{\mathrm{F}}$.
Since $q = e(N_{\mathrm{e}} -N_{\mathrm{h}})$,
one may naively expect that $N_{\mathrm{ex}}$ is minimised for a given
current when $N_{\mathrm{h}}$ or $N_{\mathrm{e}}$ vanish.


This naive expectation is correct, and it is possible to find a
time-dependent field that excites exactly $n$ electrons above the Fermi
level leaving no other disturbance in the system.
The excitation number $N_{\mathrm{ex}}$, introduced
in Eq.~(\ref{eq:charge-nex}), can be linked to noise (see below), 
thus $N_{\mathrm{ex}}$ can be measured
by sending the excited pulse on a beamsplitter (point contact)
and detecting scattering noise.


The many-body states for these pulses are found below to 
have a simple direct product form.
Considering the right moving electrons:
\begin{equation}
  \label{eq:state}
  \left| \psi \right>
  =\prod_{k=1}^n A^{\dagger}_k
  \left| 0 \right>
  , \quad
  A^{\dagger}_k
  =
  \sqrt{2\tau_k}
  \sum_{\epsilon > \epsilon_{\mathrm{F}}} 
  e^{-\xi_k \epsilon/\hbar} a^{\dagger}_\epsilon,
\end{equation}
where $\left| 0 \right>$ is the undisturbed Fermi sea, and 
\[
\xi_k = \tau_k -i t_k
,\quad
\tau_k>0
,\quad
k=1,...,n,
\]
are complex parameters specifying each pulse width and the creation moment.
Thus, a single operator $A^{\dagger}_k$ creates an electron in a
superposition of single particle eigenstates with $k\approx k_F$.
%
%
The form in Eq.~(\ref{eq:state}) applies both to a
dispersionless system, $\epsilon = v_{F} k$, and to the more general
case 
of a one to one relation between $|k|$ and
$\epsilon$.

The product form and the absence of holes
at $\epsilon<\epsilon_{\mathrm{F}}$ in (\ref{eq:state})
means that the excited particles 
due to different operators $A^{\dagger}_k$
are not entangled with one another, and that the excitation 
leaves the Fermi sea intact.
Each of the particles (\ref{eq:state}) has an exponential energy
distribution, $p(\epsilon)\propto e^{-\beta_k\epsilon}$,
of the width determined by inverse pulse duration: $\beta_k=2\tau_k/\hbar$.
From our analysis it follows that this is the only kind of energy distribution
possible under the requirement that the Fermi sea remains undisturbed.

The remarkable and somewhat paradoxical 
property of the states (\ref{eq:state}) is 
``charge imbalance,''
i.e.\ $n$ particles above Fermi level with, apparently, 
no accompanying holes. It will be seen below that such states
can nevertheless be created by electro-magnetic pulses in 
a realistic experimental situation. The accompanying holes
in fact
do appear near the Fermi level, however, at the point
$-k_{\mathrm{F}}$ opposite to where electrons are created.
A large momentum transfer $2n\hbar k_{\mathrm{F}}$ associated with 
an excitation which is slow on the scale of $\epsilon_{\mathrm{F}}$ can be 
understood as a result
of collective response of the entire Fermi sea.


It is instructive to consider the real space profile of these
states.
%
Writing $\psi(x,t) = \psi(x - v_{\mathrm{F}} t)$, 
for the case of a dispersionless Fermi system,
the real space
profile is given by the Fourier transform of the energy
representation, Eq.~(\ref{eq:state}).
For $n=1$ this gives
%
\begin{equation}
  \label{eq:2}
  \psi(x,t) = \sqrt{\frac{v_{\mathrm{F}}}{2\pi}} 
  \frac{i \sqrt{2 \tau_1} }{x - v_{\mathrm{F}} (t-t_1) + i v_{\mathrm{F}} \tau_1},
\end{equation}
i.e.\ the final many body state contains one extra electron in a state
with a Lorentzian density profile, matching the time dependence of the voltage
pulse, as discussed below. 
Similarly, there is a counter-propagating opposite sign (hole) pulse,
as illustrated in Fig.~\ref{fig:real-space}.


The application of a time dependent voltage
across a short interval (Fig.~\ref{fig:real-space}), for dispersionless
Fermi system and right-moving particles with
$k\approx k_{\mathrm{F}}$,
can be written in terms of the single-particle Hamiltonian,
\begin{equation}
  \label{eq:3}
  H = - i \hbar v_{\mathrm{F}} \partial_x + \delta(x) \hbar v_{\mathrm{F}} \phi(t)
,\quad
\frac{d\phi}{dt}=-\frac{e}{\hbar}V(t)
.
\end{equation}
Thus, the single particle states either side of $x=0$ are related by
time dependent forward scattering phase,
$\psi(x^+,t)=\psi(x^-,t)e^{i\phi(t -  x/v_{\mathrm{F}})}$.
[The intantaneous scattering approximation, expressed by
the Hamiltonian (\ref{eq:3}), means the
time of transit through the ac field region is
short compared to the pulse width $\tau$.]
The forward scattering phase $e^{i\phi(t)}$,  describing the
effect of $V(t)$ on single particle wavefunctions, defines
a canonical transformation $U$ of fermion operators 
in the many-body problem:
\be\label{eq:Ucanonical}
a_\epsilon =\sum_{\epsilon'}\la\epsilon|U|\epsilon'\ra\tilde a_{\epsilon'}
,\ 
\la\epsilon|U|\epsilon'\ra=\int e^{i(\epsilon - \epsilon^{\prime}) t/\hbar}e^{i\phi(t)}dt
.
\ee
Substituting in Eq.(\ref{eq:charge-nex}) and averaging over
the Fermi vacuum, we find the excited electron and hole numbers
\[
N_{\mathrm{e}}=\sum_{\epsilon>\epsilon_{\mathrm{F}},\epsilon'<\epsilon_{\mathrm{F}},}
|\la\epsilon|U|\epsilon'\ra|^2
,\quad
N_{\mathrm{h}}=\sum_{\epsilon<\epsilon_{\mathrm{F}},\epsilon'>\epsilon_{\mathrm{F}},}
|\la\epsilon|U|\epsilon'\ra|^2
.
\]
This means that the clean pulse condition, $N_{\mathrm{h}}=0$,
is expressed mathematically as Fourier harmonics 
$\int e^{i\omega t}e^{i\phi(t)}dt$
vanishing at $\omega<0$, where 
$\hbar \omega = \epsilon - \epsilon^{\prime}$.

%
Let us now show that clean states can be obtained by applying
a sum of Lorentzian pulses of quantized area:
\begin{equation}
  \label{eq:phase}
  e^{i \phi(t)} = 
  \prod_{k=1}^{n} \frac{t +i\xi_k^{\ast}}{t -i\xi_k}
,\quad
 V(t) =  \frac{\hbar}{e}\sum_{k=1}^n \frac{- 2\tau_k}{(t-t_k)^2 + \tau_k^2}.
\end{equation}
By Fourier transforming $e^{i\phi(t)}$ which is analytic in the lower 
halfplane of complex $t$ we see that the negative Fourier harmonics 
vanish. For example, at $n=1$,
\begin{equation}
  \label{eq:phase_ft}
\la\epsilon| U|\epsilon'\ra = 
\int \frac{dt}{2\pi}  
\frac{t +i\xi_1^{\ast}}{t - i\xi_1} e^{i\omega t}
  =
  \delta(\omega) - 2 \tau_1 e^{- \omega \xi_1}\theta(\omega).
\end{equation}
Thus we have $N_{\mathrm{h}}=0$ and $N_{\mathrm{e}}=n$, proving that the time dependence (\ref{eq:phase}) indeed
leads to clean pulses.
Sign reversal in (\ref{eq:phase}),
$\phi,V\to -\phi,-V$, 
gives pulses which create $n$ holes in a similar clean fashion.

%
%
%

The form of Eq.\,(\ref{eq:phase}) is suggested
by recalling
several previous instances when Lorentzian $V(t)$ of quantized area
appeared, such as tunneling
of phase in Josephson junctions~\cite{Korshunov87}, 
charge pumping noise~\cite{HWLee0503},
quenching of Coulomb blockade~\cite{Nazarov99},
1d quantum hydrodynamics~\cite{Wiegmann06}, and vertex operators
in Quantum Hall systems~\cite{Stone91}.
Besides producing particularly low scattering noise~\cite{HWLee0503},
the pulses (\ref{eq:phase}) were found to give rise to
strikingly simple counting statistics~\cite{ivanov97,ambrumenil05}. 
The reason for the latter, which previously was unknown, will
be clarified below.
%


The clean states obtained by
applying an electric field with the time
dependence (\ref{eq:phase}) 
are of the form described in Eq.~(\ref{eq:state}).
We first show it in the case $n=1$ (total phase increase $2\pi$),
for simplicity setting $\hbar=1$.
%
%
The initial state, i.e.\ the filled Fermi sea,
is associated with a projection $\hat n$ 
on the states with $\epsilon<\epsilon_{\mathrm{F}}$ in the single particle Hilbert space. 
After
applying a unitary evolution $U$, the evolved Fermi sea 
will be associated with
$\hat n'=U\hat nU^{\dag}$. 
The particles taken from below to above the Fermi
surface are thus associated with $U_{+-}=(1-\hat n)U\hat n$, which
in the current case, 
described by Eq.(\ref{eq:phase_ft}), has the form:
$\la\epsilon|U_{+-}|\epsilon'\ra=-2\tau_1e^{- (\epsilon-\epsilon') \xi_1}=
- 2 \tau_1 e^{- \epsilon \xi_1}e^{ \epsilon' \xi_1}$.
Owing to multiplicativity of exponential,
$U_{+-}$
is a rank one matrix, and
as will be shown next, this means that only a single
particle is excited, ensuring the minimal excitation property. 




The requirement that the matrix is of rank one, which is basis
independent, requires $U_{+-}$ to be of the form
\be\label{U_odd}
U_{+-}=c|\phi_{+}\ra \la \phi_{-}|,
\ee
with the states $\phi_{-}$, $\phi_{+}$ inside and outside
the Fermi sea.
%
This structure implies there
can be at most one particle excited above the Fermi surface.
Consider a hypothetical 
transition in which two or more
particles are excited from levels below to above the Fermi surface.
For any given pair of initial levels below the Fermi surface, $a,b$,
and  final levels above the Fermi surface, $a',b'$, there are two
ways such transition
can be achieved; $a \to a', b \to b'$ and $a \to b', b \to a'$.
However, since 
$U_{k\to k'}=\la k'|U_{+-}|k\ra =c \la k'|\phi_{+}\ra \la \phi_{-}|k\ra$
for any pair of states $|k\ra$, $|k'\ra$
it follows that the two-fermion transition amplitude vanishes:
\be\label{eq:no2particles}
U_{a \to a'} U_{b \to b'} - U_{a \to b'} U_{b \to a'}=0.
\ee
Thus the Fermi statistics blocks two-particle transitions.
Similarly, the requirement for having no more than $n$
particles/holes excited is that $U_{+-}$ is a matrix of
rank $n$.


The $\delta(\omega)$ term in Eq.~(\ref{eq:phase_ft}) might 
suggest
there is a finite probability that the pulses (\ref{eq:phase}) 
produce no excitation. 
However 
the corresponding weight
 is exponentially small in the pulse width,
$\tau \epsilon_{\mathrm{F}}/\hbar\gg 1$, and thus can be ignored.
Thus,  a single
particle is created in the state $\left|\phi_{+}\right>$, and by matching
the form of $U_{+-}$ in Eq.~(\ref{eq:phase_ft}) and Eq.~(\ref{U_odd}),
one finds that the state created $\left|\phi_{+}\right>$ is
that of Eq.~(\ref{eq:state}).
Since $U_{-+}=0$, it is clear 
 there are no holes created moving to the right.
In contrast, the left moving particles have $U_{+-}=0$ and $U_{-+}$
of rank one. 
Thus there is one hole created near $-k_{\mathrm{F}}$ in momentum space,
but no particles excited.


Having understood the single pulse, one
can now consider combining such pulses.
Consider two pulses, 
%
\begin{equation}
  \label{eq:double-pulse-defn}
  e^{i\phi(t)} = 
  \left( 
    \frac{t +i \xi_1^{\ast}}{t - i\xi_1}
  \right)
  \left( 
    \frac{t +i \xi_2^{\ast}}{t - i\xi_2}
  \right).
\end{equation}
Using the result of a single pulse acting on the vacuum state,
and introducing $\tilde A^{\dagger}_2 = U_1 A^{\dagger}_2 U_1^{\dagger}$,
(i.e.  transforming each single particle operator in $A^{\dagger}_2$
by the unitary matrix $U_1$ as in Eq.(\ref{eq:Ucanonical}))
we can write the result of two pulses
as:
%
\be
  \label{eq:double-pulse-reorder}
  U_1 U_2
  \left| 0 \right>
  = 
  U_1
  A^{\dagger}_2 \left| 0 \right>
  =  
  U_1
  A^{\dagger}_2 
  U_1^{\dagger}U_1
  \left| 0 \right>
  = \tilde A^{\dagger}_2 A^{\dagger}_1 \left| 0 \right>.
\ee
%
%
Using the matrix elements, Eq.~(\ref{eq:phase_ft}),
we find
\begin{eqnarray}
  \label{eq:transform-A}
\tilde A^{\dagger}_2 &=&  A^{\dagger}_2 
  -
  2 \tau_1 \sqrt{2 \tau_2}
  \int_0^{\infty} \!\!\!\!\! d\nu
  \int_0^{\infty} \!\!\!\!\! d\omega
  e^{-\omega \xi_1 - \nu \xi_2} a^{\dagger}(\omega + \nu),
  \nonumber\\
  &=&
  A^{\dagger}_2 
  - 
  \frac{2 \tau_1}{\xi_1-\xi_2} \left(
      A^{\dagger}_2 - \sqrt{\tau_2/\tau_1} A^{\dagger}_1
  \right).
\end{eqnarray}
Since the operators $A^{\dagger}$ are fermionic, $A^{\dagger 2}_1=0$,
and so the result of two phase pulses is given by:
\begin{equation}
  \label{eq:double-pulse-result}
  U_1 U_2
  \left| 0 \right>
  =
  \frac{
        \xi_1^\ast + \xi_2
  }
  {
        \xi_2-\xi_1
  }
  A^{\dagger}_2  A^{\dagger}_1
 \left| 0 \right>.
\end{equation}
Similarly, for $n$ pulses, Eq.~(\ref{eq:phase}), we obtain the algebra
\be\label{eq:Laughlin}
\prod_{k=1}^n 
U_k
|0\ra =
\prod_{k>k'}\frac{\xi_{k'}^\ast + \xi_k}{\xi_{k'}-\xi_k}
A^{\dagger}_n A^{\dagger}_{n-1} \ldots A^{\dagger}_1
 \left| 0 \right>.
\ee
The expression for holes near $-k_{\mathrm{F}}$ is similar,
but with the complex conjugate of the factor multiplying the operators
$A^{\dagger}_k$.
When the creation times are equal, $t_k = t_{k'}$, the prefactor in
Eq.~(\ref{eq:Laughlin}) depends only on $\tau_k$, is real
and antisymmetric in permutations. 
This means that even looking at just the
Fermi point $k_{\mathrm{F}}$, the pulse creation operators effectively commute
(since $A^{\dagger}_k$ anticommute). Remarkably,
the form of the expression (\ref{eq:Laughlin}) 
is similar to the Laughlin state in a
Quantum Hall system.

In the limit $\xi_k \to \xi_{k'}$
the prefactor in Eq.~(\ref{eq:Laughlin})
appears to diverge, however the product of two identical single
fermion operators would vanish. Taking these limits together, the
result is expressed through $A^{\dagger}_k\partial_{\xi_k}A^{\dagger}_k$,
i.e.\ is non-zero, but 
not of the
form of Eq.~(\ref{eq:state}).

%

Similarly, a pulse and an anti-pulse give:
%
%
%
\begin{equation}
  \label{eq:puls-anti-pulse}
  e^{i\phi(t)} = 
  e^{-i\phi_1(t)}
  e^{i \phi_2(t)}
  =
  \left( 
    \frac{t - i\xi_1}{t +i\xi_1^{\ast}}
  \right)
  \left( 
    \frac{t +i\xi_2^{\ast}}{t - i\xi_2 }
  \right),
\end{equation}
with $\tau_1, \tau_2 > 0$.
This combination corresponds to zero total phase change, and thus 
the associated pulse carries zero net current. 
Such a pulse 
exhibits a higher degree of particle-hole entanglement
than that of Eq.~(\ref{eq:state}).

%

Introducing operators $B$, creating a single hole below the
Fermi level near $k_{\mathrm{F}}$, to describe the result of $e^{-i \phi_1(t)}$, one
can follow the same procedure as in
Eqs.~(\ref{eq:double-pulse-reorder}),(\ref{eq:transform-A}):
%
\bea
  \label{eq:1}
&&    \bar{U}_1 U_2
    \left| 0 \right>
    =
    \left( 
      \bar{U}_1 
      A^{\dagger}_2 
      \bar{U}_1^{\dagger} 
    \right) 
B^{\nodagger}_1 \left| 0 \right>
 \nonumber\\
&& = \frac{\xi_2-\xi_1}{\xi_2+\xi_1^\ast}A^{\dagger}_2 B^{\nodagger}_1
\left| 0 \right>
-\frac{2 \sqrt{\tau_1 \tau_2}}{\xi_2+\xi_1^\ast}\left| 0 \right>.
\eea
%
where as in Eq.~(\ref{eq:transform-A}),
$\bar{U}_1 A^{\dagger}_2 \bar{U}_{1}^{\dagger}$
has the form of $u A^{\dagger}_2+v B^{\dagger}_1$,
with the last term, applied to the state $B^{\nodagger}_1 \left| 0 \right>$,
producing $\left| 0 \right>$.
Hence the final state is a superposition
containing two parts:
an electron-hole pair
$A^{\dagger}_2B^{\nodagger}_1$, with an extra electron just above the Fermi
level, and a hole just below, and a part which is the unperturbed
Fermi sea.
As 
$\tau_1,t_1 \to \tau_2,t_2$ 
the weight of the excited
part decreases, and the state approaches the unperturbed state.


It is natural and interesting to generalize the above results to Luttinger 
liquids, where charge fractionalization~\cite{Kane94}
allows one to create clean pulses carrying noninteger charge.
Here we consider chiral Luttinger liquid of the kind realized
on a Quantum Hall edge. In the simplest case of a single 
chiral mode the Largrangian~\cite{Wen90} for displacement
field 
coupled to an ac voltage is of the form
\be\label{eq:LagrangianLL}
\L =\int \lp
\frac1{4\pi}\p_x\theta (\p_t-v\p_x)\theta +\frac{\sqrt{\nu}}{2\pi}\p_x\theta
eV(t,x)\rp dx
.
\ee
%
The operators $\psi(t,x)\propto e^{i\sqrt{\nu}\theta(t,x)}$
($\nu=1/m$ for Laughlin state) describe quasiparticles of charge 
$e_\ast=\nu e$ which obey Fermi statistics~\cite{Wen90}. 
Solving for the displacement operator time dependence,
$(\p_t-v\p_x)\theta=-\sqrt{\nu}eV(t,x)$,
we obtain the Heisenberg evolution
of a quasiparticle: 

\be\label{eq:phaseLL}
\psi(t,x)=\psi(0,x+vt)\exp\lp -i\int_0^t e_\ast V(t',x')dt'\rp
,
\ee
$x'=x+v(t-t')$.
This means that the phase picked up 
after passing through the ac field 
is $\delta\phi=-\int e_\ast V(t')dt'$ (at fast passage, i.e.\ 
when $\tau v\gg L$, 
where $L$ is the width of the region where field is applied).
From Eq.~(\ref{eq:phaseLL}) it is clear that the problem 
maps exactly onto the free fermion problem studied above,
with electrons replaced by fractional charge quasiparticles. The
optimal pulses (phase increase $2\pi$) are Lorentzians
of area $\int Vdt=h/\nu e$,
with the corresponding excitation carrying charge $e_\ast$.


To discuss the possibility of testing the above predictions 
experimentally, let us link the excitation number with noise.
The operator counting
particle-hole excitations,
\begin{math}
  \hat{N}_{\mathrm{ex}} = 
  \sum_{\epsilon < \epsilon_{F}} 
  a^{\nodagger}_{\epsilon}  a^{\dagger}_{\epsilon}
  +
  \sum_{\epsilon > \epsilon_{F}} 
  a^{\dagger}_{\epsilon}  a^{\nodagger}_{\epsilon}
\end{math},
has an expectation value which
can be found from the trace of $\hat{N}_{\mathrm{ex}}$ 
with the single particle density
matrix describing the system after the field is applied.
Writing this trace in the time domain, the field leads to a
time-dependent phase, giving:
\begin{eqnarray}
  \label{eq:N_ex}
  \langle \hat{N}_{\mathrm{ex}} \rangle
  &=&
  \int dt \int dt^{\prime}
  N_{\mathrm{ex}}(t,t^{\prime}) \left( 
    e^{-i\phi(t^{\prime})}
    \rho_0(t^{\prime},t)
    e^{i\phi(t)}
  \right)
  \nonumber\\
  &=&
\sum_{\pm}  \int \frac{dt}{2\pi} \int \frac{dt^{\prime}}{2\pi}
  \frac{
    \pi^2 T^2 e^{i \left( \phi(t) -\phi(t^{\prime}) \right) } 
  }{
    \sinh^2\pi T \left(t - t^{\prime} - i 0^{\pm} \right)
}
  \nonumber\\
  &=&
\int \frac{d\omega}{4\pi^2}
\left|\int e^{i\phi(t)+i\omega t} dt\right|^2 \omega \coth\frac{\omega}{2T}
  ,
\end{eqnarray}
where the average is taken over a state with temperature $T$.
This expression is equal, up to a system-dependent factor, to the noise
created as a result of the pulse scattering on 
a barrier~\cite{HWLee0503}. Thus one can measure
$N_{\mathrm{ex}}$ directly by passing the excited pulse through a beamsplitter,
such as point contact, and detecting resulting current noise.

The dependence (\ref{eq:N_ex}) 
on $\phi(t)$
is such that for the non-quantized pulses
(phase increase not a multiple of $2\pi$) $N_{\mathrm{ex}}$ exhibits a log
divergence~\cite{HWLee0503}. The quantized pulses,
characterized by smaller $N_{\mathrm{ex}}$, produce lower noise.
Among those the Lorentzian pulses (\ref{eq:phase})
provide absolute minimum to the noise.

The noise-free character of the states (\ref{eq:state}), comprised
of a few particles and undisturbed
Fermi sea, makes the effect of their scattering
very simple to interpret. 
If no dc voltage is present, i.e.\ the 
Fermi level is the same in all reservoirs, as in the above discussion,
scattering of the states (\ref{eq:state})
will leave the filled Fermi sea intact, and the only
contribution will arise from the excited particles. The noise
in such a case, as well as the entire counting statistics, should be of 
single particle character. This property of Lorentzian voltage pulses
applied to the barrier was noted in Refs.\cite{HWLee0503,ivanov97}.

To summarize, we described a realization of a source of 
single fermions which does not rely on electron confinement.
The particles can 
be excited directly out of a Fermi sea by a carefully designed perturbation
without creating additional noise. 
The profile of pulses required for that is Lorentzian,
with quantized area and duration 
shorter than $\hbar/kT$ and
long compared to $\hbar/\epsilon_{\mathrm{F}}$.
For typical values $T=10\,{\rm mK}$ and $\epsilon_{\mathrm{F}}=10\,{\rm meV}$
this gives a realistic frequency range $200\,{\rm MHz}< \nu < 2\,{\rm THz}$.
In a Luttinger liquid this provides a source of quasiparticles with
fractional charge.
The excited particles propagate ballistically
and can be used to transmit quantum states across the system, create 
entangled particle-hole pairs, and analyze them by noise measurement. 

\begin{acknowledgments}
  This work was supported by NSF-NIRT DMR-0304019; J.K. acknowledges
  financial support from the Lindemann Trust.
\end{acknowledgments}


\end{document}